\documentstyle[amssymb,preprint,aps]{revtex}

\begin{document}
\title{A Qubit on the Faraday Mirror}
\author{F. De Martini, F. Sciarrino and V. Secondi}
\address{Dipartimento di Fisica and \\
Istituto Nazionale per la Fisica della Materia\\
Universit\`{a} di Roma ''La Sapienza'', Roma, 00185 - Italy}
\maketitle

\begin{abstract}
In the context of Quantum Information (QI)\ the ''Faraday Mirror'' acts as a 
{\it non-universal} NOT Gate. As such its behaviour complies with the
principles of quantum mechanics. This non trivial result, at the core of
some recent misinterpretations in the QI\ community, has been reached by a
thorough experimental investigation of the properties of the device
including the adoption of modern {\it Quantum Process Tomography.} In
addition, the ''universal optical compensation'' method devised by Mario
Martinelli, of common use in long distance quantum-cryptography, has been
fully investigated theoretically and experimentally.
\end{abstract}


The today common interest in the ''optimal'' realization of ''forbidden''
quantum information processes and the recent realizations of the Universal
Not Gate \cite{1,2,3} have stressed the somewhat intriguing role of the
''Faraday Mirror'' $(FRM)$\ in connection with the ''spin flipping''
process, i.e. the {\it positive} map (P-map) that should transform any qubit 
$\left| \Psi \right\rangle $ into the orthogonal one $\left| \Psi _{\perp
}\right\rangle $, i.e. such as $\left\langle \Psi \mid \Psi _{\perp
}\right\rangle =0$ \cite{4}. In other words, this map should draw any point $%
P$ representative of $\left| \Psi \right\rangle $ on the Poincare' sphere
into its ''antipode'' $P\prime $ representative of $\left| \Psi _{\perp
}\right\rangle $. In this connection the $FRM$ issue\ has been taken up by
an extended comment in a recent review paper on Quantum Cryptography \cite{5}%
. In view of some recent misinterpretations of the phenomenon we do believe
that this matter requires a necessary clarification \cite{5}.

The $FRM$\ is a simple, somewhat exotic device consisting of an optical
mirror $(M)$ that reflects a monochromatic light beam, associated with a
(single gaussian) mode ${\bf k}${\bf ,} into the mode $-{\bf k}$. In front
of the mirror is placed a ''Faraday Rotator'' $(FR)$, a laboratory device
generally used to avoid spurious laser back-reflection effects, in which a
transparent paramagnetic spin glass rod is acted upon by a static magnetic
field $\overrightarrow{B}$, parallel to ${\bf k}$. By the magneto-optical
Faraday effect the incoming polarization $(\pi )$ is ''rotated'' by keeping
the same $\pi -$character: a ''linear'' $\pi $ is transformed into a
''linear'' $\pi ^{\prime }$, a ''circular'' $\pi $ into a ''circular'' $\pi
^{\prime }$ etc. In particular, the field $\overrightarrow{B}$ and the $\pi
- $rotation ''angle'' $\theta $ can be ''tuned'' as to realize the
orthogonality between $\pi $ and $\pi ^{\prime }$\cite{6}. Indeed, precisely
for this useful property the $FRM$\ device is commercially advertized in
optics and in laser technology. Furthermore, according to an original
proposal by Mario Martinelli \cite{7}, it can also be adopted to eliminate
spurious birefringence effects in fiber optics technology, as we shall see.

From a quantum mechanical perspective this process, as it is commonly
understood, may indeed represent a real puzzle. The obvious point of concern
is precisely his alleged property of realizing a {\it universal} spin-flip
transformation $\left| \Psi \right\rangle \rightarrow \left| \Psi _{\perp
}\right\rangle $ in spite of this map being ''forbidden'' by quantum
mechanics $(QM).$ Indeed according to $QM$ only the {\it complete-positive}
maps (CP-maps)\ are realizable by Nature. In addition, since the $FRM$ is a 
{\it passive}, and then noiseless device, that transformation should be {\it %
exact}, i.e. implying a {\it full transfer} of quantum information from $%
\left| \Psi \right\rangle $ to $\left| \Psi _{\perp }\right\rangle $. In
order to avoid further speculations, the present paper intends to clarify
the problem by a clear-cut recourse to the experiment.

A Spontaneous Parametric Down Conversion (SPDC) process was excited in a 1.5
mm thick BBO ($\beta -$barium borate) nonlinear (NL)\ crystal slab, by a
mode-locked pulsed UV beam with wavelength (wl) $\lambda _{p}$:\ Figure 1.
By this setup entangled photon pairs in a {\it singlet-state} of
polarization $(\pi )$ were created with common wl's $\lambda =%
{\frac12}%
\lambda _{p}=795nm$. Each excitation laser pulse had a time duration $\tau
\approx $140 f$\sec $. Because of the low pump intensity, the probability of
the unwanted simultaneous $N=2$ pairs generation in the NL\ crystal was
found $\approx $ $10^{-2}$ smaller than the one for a single pair: $N=1$. In
virtue of the nonlocal correlation existing between the photons of each
pair, the qubit $\left| \Psi \right\rangle $ injected over the mode ${\bf k}%
_{2}$ into the $FRM$ device was deterministically selected by the
measurement apparatus coupled to mode ${\bf k}_{1}$. This\ one consisted of
a $\pi -$analyzer, i.e. a polarizing beam-splitter $(PBS)$, a pair of $%
(\lambda /2+\lambda /4)$ optical wave-plates (wpl) and of a tunable Babinet
Compensator $(BC)$. The detector $D_{1}$, and the end detector $D_{2}$, were
equal SPCM-AQR14 Si-avalanche single photon units with quantum efficiencies $%
QE^{\prime }s\cong 0.55$. One interference filter with bandwidth $\Delta
\lambda =3nm$ was placed in front of each $D$. The Faraday rotator $(FR)$, a
CVI-FR-660/1100-8\ device accurately $\overrightarrow{B}-$tuned for the wl $%
\lambda $, could be optionally removed from the optical network. It was
terminated by a 100\% reflectivity mirror $M$. In order to avoid the use of
other mirrors, the output qubit $\left| \Psi ^{\prime }\right\rangle $
emitted by $FRM$ over the mode $-{\bf k}_{2}$ was transmitted through the
transparent BBO NL\ slab and finally detected by the end measurement
apparatus, made of\ a $PBS$, a pair of $(\lambda /2+\lambda /4)$ wpl set and
a detector $D_{2}$. Three quartz slabs $(Q)$ inserted on the modes ${\bf k}%
_{1},$ ${\bf k}_{2},$ $-{\bf k}_{2}$ and the Babinet compensator on mode $%
{\bf k}_{1}$ secured the accurate compensation of all residual birefringence
effects, e.g. of the BBO slab.

In a first experiment we wanted to fully characterize by standard quantum
tomography the input entangled state generated by SPDC,\ {\it in absence} of 
$FRM$ \cite{8}. The reconstructed density matrix $\rho _{k1,k2}^{in}$ of the
photon pair is reported in Fig. 2-{\bf a}. There the realization of the {\it %
singlet} input pure state condition $\left| \Phi \right\rangle
=2^{-1/2}\left( \left| H\right\rangle _{1}\left| V\right\rangle _{2}-\left|
V\right\rangle _{1}\left| H\right\rangle _{2}\right) $ is clearly shown,
where $\left| H\right\rangle _{i}$ and $\left| V\right\rangle _{i}$ stands
for {\it linear}-$\pi $ horizontal and vertical qubits, associated
respectively to modes ${\bf k}_{i}$, $i=1,2$. We then analyzed the
transformation induced by the mirror reflection only (a condition henceforth
referred to as: $M-${\it reflection}) and by the complete $FRM$ device ($%
FRM- ${\it reflection)}. By the knowledge of $\left| \Phi \right\rangle $ we
prepared the input qubit $\left| \Psi \right\rangle $ injected over ${\bf k}%
_{2}$ by exploiting the nonlocal $\pi -$correlation existing between the two
photons of each pair. Precisely, we limited ourselves to prepare six input
qubits $\left| \Psi \right\rangle :$ $\left| H\right\rangle $, $\left|
V\right\rangle $, $\left| L_{\pm }\right\rangle \equiv 2^{-%
{\frac12}%
}(\left| H\right\rangle \pm \left| V\right\rangle )$, $\left| C_{\pm
}\right\rangle \equiv 2^{-%
{\frac12}%
}(\left| H\right\rangle \pm i\left| V\right\rangle )$. The results of two
experiments carried out respectively in absence and in absence of $FR$, were
the following:

$M-${\it reflection}: $\left| H\right\rangle $ $\rightarrow \left|
H\right\rangle $; $\left| V\right\rangle \rightarrow \left| V\right\rangle
\, $; $\left| L_{\pm }\right\rangle \rightarrow \left| L_{\mp }\right\rangle 
$; $\left| C_{\pm }\right\rangle \rightarrow \left| C_{\mp }\right\rangle $,
with fidelities ${\cal F}$ = $0.98$, $0.98$, $0.90$, $0.90$ respectively.

$FRM-${\it reflection}: $\left| H\right\rangle $ $\rightarrow \left|
V\right\rangle $; $\left| V\right\rangle \rightarrow \left| H\right\rangle $%
; $\left| L_{\pm }\right\rangle \rightarrow \left| L_{\pm }\right\rangle $; $%
\left| C_{\pm }\right\rangle \rightarrow \left| C_{\mp }\right\rangle $ with 
${\cal F}$ =$0.99$, $0.99$, $0.92$, $0.92$ respectively.

Here the arrows indicate the mapping from the input to the output states,
associated with the input/output modes ${\bf k}_{2}$ and $-{\bf k}_{2}$,
correspondingly. The ${\cal F}$ values, here defined as the statistical
fractions of the ''right'' outcomes, were limited by an imperfect correction
of the optical walk-off inside the NL crystal. We can see that, in agreement
with $QM$, the $FRM\;$device {\it does not }realize the {\it universal}
spin-flip action but rather a {\it CP-map}, i.e. a {\it unitary}
transformation expressible by a definite rotation around the center of the
Poincare' sphere of the unit vector $\overrightarrow{p}$. This vector is
then drawn by the transformation into a position $\overrightarrow{p}^{\prime
}$ in the tri-dimensional {\it spin space} with orthogonal axes $\sigma _{1}$%
, $\sigma _{2}$, $\sigma _{3}$:\ Figure 1, inset. Correspondingly, the
vertex of $\overrightarrow{p}$, i.e. the point $P$ expressing the input $%
\left| \Psi \right\rangle $ on the surface of the Poincare' sphere is drawn
into the point $P^{\prime }$ expressing the output $\left| \Psi ^{\prime
}\right\rangle $.

Let's gain insight into the overall process by following the path through $%
FRM$ of the short light pulse associated with any single photon carrying the
qubit. Consider the (real) spatial reference frame (x,y,z) the qubit is
referred to before reflection by $M$:\ let the axes x, y belong to the plane
of $M$, while the z-axis, orthogonal to $M$, is parallel to and oriented as
the input photon momentum $\hslash {\bf k}_{2}$. Assume that the qubit's $%
\pi -$states are identified according to the {\it right-hand} (r.h.)
coordinate convention, henceforth referred to as ''{\it \ frame-convention}%
''. Of course, the preservation of the {\it \ frame-convention} upon $M-$ or 
$FRM-${\it reflection }is a necessary requirement for a correct
understanding of the reflection dynamics, since the input and output qubits
are identified precisely on the basis of the real-space reference frame.
This implies that, after reflection the photon's $\pi -$states of any photon
are identified on the basis of a new coordinate frame (x'=x, y'=-y, z'=-z),
obtained by rotating the old frame around the $\overrightarrow{x}$ axis by
an angle $180%
{{}^\circ}%
$. This easily explains the (apparent) puzzling behaviour with respect to
state-orthogonality of the input/output {\it linear polarization} states
under $FRM-${\it reflection}.

In order to visualize the process with the help of the Poincare' sphere, let
us consider the $M-${\it reflection} first. Let the point $P$ to represents
the input qubit when the input photon hits $M$, as said: $\left| \Psi
\right\rangle =\psi _{1}\left| H\right\rangle $ $+\psi _{2}$\ $\left|
V\right\rangle $, $\left| \psi _{1}\right| ^{2}+$ $\left| \psi _{2}\right|
^{2}=1$, being $\psi _{i}$ the complex components of the spinor $%
{\psi _{1} \atopwithdelims|| \psi _{2}}%
$. Classical optics and our experimental results above show that the
reversal of the particle motion and the preservation of the {\it %
frame-convention} upon $M-${\it reflection} implies a rotation of $%
\overrightarrow{p}$ \ around the spin reference axis $\sigma _{3}$ by an
angle $180%
{{}^\circ}%
$, i.e. by the unitary map: ${\Bbb U}_{3}=\exp (i%
{\frac12}%
\pi \sigma _{3})=i\sigma _{3}$. Note that, after the corresponding
representative point $P$ being drawn into $P^{\prime }$ by the $M-${\it %
reflection }transformation{\it , }the {\it same} Poincare' sphere can still
be adopted for the future qubit dynamics, indeed a useful condition when
searching for a\ possible ''antipode'' of $P$ over the sphere. In summary,
the $M-${\it reflection} process transforms the input qubit $\left| \Psi
\right\rangle $ above as follows: ${\Bbb U}_{3}%
{\psi _{1} \atopwithdelims|| \psi _{2}}%
=i%
{\psi _{1} \atopwithdelims|| -\psi _{2}}%
$\cite{9}.

Let us turn now our attention to the $FRM-${\it reflection}. If \ $P$ now
represents the qubit before entering the device, the first passage through $%
FR$ draws $P$ into $O$ by a positive rotation by $90%
{{}^\circ}%
$ around $\sigma _{2}$, i.e. corresponding to the evolution operator: ${\Bbb %
U}_{2+}=\exp (i%
{\frac14}%
\pi \sigma _{2})$. The $M-${\it reflection} contributes with the map ${\Bbb U%
}_{3}$, leading from the point $O$ to $O^{\prime }$, as just seen. The
second passage through $FR$ after $M-${\it reflection} is represented by $%
{\Bbb U}_{2-}=\exp (-i%
{\frac14}%
\pi \sigma _{2})$ since the change of the spatial reference frame leaves
unchanged the (axial) magnetic-polarization pseudo-vector $\overrightarrow{%
{\cal M}}{\cal \varpropto }\overrightarrow{B}$ while the sign of the phase $%
\pi /2$ is inverted because of the particle's propagation reversal \cite{10}%
. In summary, the overall evolution of\ the input qubit $\left| \Psi
\right\rangle $ through the $FRM$ leaves it in the final state: $\left| \Psi
^{\prime }\right\rangle $=${\Bbb U}_{FRM}\left| \Psi \right\rangle $= ${\Bbb %
U}_{2-}{\Bbb U}_{3}{\Bbb U}_{2+}\left| \Psi \right\rangle $ = $%
{\frac12}%
(1-i\sigma _{2})\times (i\sigma _{3})\times (1+i\sigma _{2})\left| \Psi
\right\rangle \ $=$i\sigma _{1}\left| \Psi \right\rangle $, since: $\sigma
_{i}\sigma _{j}=i\sigma _{k}$ for an even permutation of $i,j,k$.
Correspondingly, the initial point $P$ on the sphere is drawn into the
successive positions $O$, $O^{\prime }$and $P^{\prime }$. Correspondingly,
the overall $FRM$ process results in a overall $180%
{{}^\circ}%
$ rotation around the axis $\sigma _{1}$ of the initial representative
vector $\overrightarrow{p}$, i.e. the input qubit is acted upon by the {\it %
unitary}: ${\Bbb U}_{FRM}=\exp (i%
{\frac12}%
\pi \sigma _{1})=i\sigma _{1}$, as seen. As we can see, the final
representative point $P^{\prime }$ {\it is not }the antipode of {\it any
possible} $P$ on the sphere and then $FRM\;${\it does not} realize the {\it %
universal} quantum-NOT Gate, as expected from QM. Indeed the spin-flip
action is realized exactly {\it only} for all representative points $P$
belonging to the equatorial plane orthogonal to the $\sigma _{1}$ axis, such
as for the qubits $\left| H\right\rangle ,$ $\left| V\right\rangle $ and $%
\left| C_{\pm }\right\rangle $.

As an additional deeper-level demonstration, a very general characterization
of the $FRM$ device was attained experimentally by the entanglement-assisted 
{\it Quantum Process Tomography }${\it (QPT)\ }$\cite{11}. Let us outline
the basics of\ this approach, by first representing the state of the input
qubit corresponding to a point $P$ on the Poincare' sphere, by the density
operator $\rho _{k2}=%
{\frac12}%
({\Bbb I}+\overrightarrow{r}\cdot \overrightarrow{\sigma })$. Then this
qubit can be represented by a $4-$dimensional vector $(1,\overrightarrow{r})$%
. The single map ${\cal E(\rho )},$ expressing the $FRM-${\it reflection},
is completely described by a $4\times 4$ real matrix ${\bf M}_{FRM}{\bf ,}$
which maps $\rho _{k2}$ into the density matrix: $\rho _{-k2}^{\prime }={\bf %
M}_{FRM}\rho _{k2}$ \cite{12}. A most important, exclusive property of \ the 
$QPT$\ method is the exploitation of the {\it quantum parallelism}
associated with the entanglement condition \cite{11}. Indeed, in our
experiment the matrix ${\bf M}_{FRM}$ was experimentally reconstructed with
the use of the input pure {\it entangled state} $\rho _{k1,k2}^{in}$of the
two correlated qubits. The first step consisted of performing the quantum
state tomography of that state $\rho _{k1,-k2}^{in}$(Fig.2-{\bf a-}left
side). Then by ${\it QPT}$ the qubit associated with mode ${\bf k}_{1}$,
call it ''{\it qubit 1}'', was left unchanged while the ''{\it qubit 2}'',
associated with the mode ${\bf k}_{2}$ underwent the ${\cal E(\rho )}{\cal -}
$transformation. Note that this procedure implies the investigation of the
unknown map ${\cal E(\rho )}$, i.e. of $FRM-${\it reflection,} by a {\it %
complete} span over the Hilbert space ${\cal H}_{2}$ of the injected ''qubit
2'' because of its {\it mixed-state} condition. The final state of the two
qubits $\rho _{k1,-k2}^{out}={\Bbb I}_{k1}\otimes {\cal E}_{k2}(\rho
_{k1,k2}^{in})$ was again investigated by tomography (Fig.2-{\bf a-}right
side). Finally, the matrix ${\bf M}_{FRM}$ was estimated by means of the
experimentally determined density matrices $\rho _{k1,k2}^{in}$ and $\rho
_{k1,-k2}^{out}$ by adoption of an inversion software \cite{11}. Remind the
result ${\Bbb U}_{FRM}=i\sigma _{1}$obtained by the theoretical analysis of
the $FRM-${\it reflection, }given above. We may then compare the matrix $%
{\bf M}_{FRM}$\ associated to the unknown $FRM$ map ${\cal E(\rho )}$ with
the corresponding one expressing the $\sigma _{1}$ Pauli operator: 
\begin{equation}
{\bf M}_{\sigma _{1}}=\left( 
\begin{array}{cccc}
1 & 0 & 0 & 0 \\ 
0 & 1 & 0 & 0 \\ 
0 & 0 & -1 & 0 \\ 
0 & 0 & 0 & -1
\end{array}
\right)  \label{sigmax}
\end{equation}
The overlap of two generic maps ${\cal E}$ and ${\cal L}$ is usually defined
by the fidelity: ${\cal F}\left( {\cal E},{\cal L}\right) =\int d\Psi {\cal F%
}\left[ {\cal E}\left( \left| \Psi \right\rangle \left\langle \Psi \right|
\right) ,{\cal L}\left( \left| \Psi \right\rangle \left\langle \Psi \right|
\right) \right] $\cite{13}. In the present context we obtain: ${\cal F}%
\left( \sigma _{1},{\cal E(\rho )}\right) =\frac{1}{2}+\frac{1}{3}\left(
M_{22}-M_{33}-M_{44}\right) =\left( 99.7\pm 0.3\right) \%$ where $M_{ii}$ is
the $ii-$element of the matrix ${\bf M}_{FRM}$. The $QPT$ reconstruction of $%
{\bf M}_{FRM}$ is shown in Figure 2-{\bf b). }We may compare the diagram
reported there with the structure of the matrix ${\bf M}_{\sigma _{1}}$,
Eq.1. The correspondence is quite impressive \cite{14}.

In summary, our experimental results support the previous theoretical
analysis and then the $FRM\;$''puzzle'' can be considered fully clarified.
In addition, these conclusions emphasize once again the relevance of the two
different {\it Universal}-{\it Not Gates} ($U-NOT$)\ schemes that have been
recently reported in the literature \cite{2,3}. Indeed these schemes fully
implement the {\it universality} condition, albeit {\it optimally}, i.e. by
a fidelity $F<1$. This last condition is physically achieved in the {\it %
deterministic}, {\it parametric} {\it amplification} $U-NOT$\ scheme by
addition to the output ''signal'' of a ''noise'' contribution arising from
the QED\ vacuum fluctuations \cite{2}. In the{\it \ probabilistic,} linear%
{\it \ Tele-}$U-NOT$ scheme, a partial cancellation of information is due to
the statistical, i.e. non deterministic character of the device \cite{3}.

For the sake of completeness, let's investigate here also the interesting ''$%
FRM$ -compensation'' process discovered by Mario Martinelli that has
recently found a widespread application whenever an active compensation of
optical path fluctuations is required, as in long distance quantum
cryptography applications with optical fibers \cite{5,7}. The optical scheme
of the process is the following. Suppose that Alice (A)\ and Bob (B), two
receiving/sending QI\ stations are connected by an optical device ($PC$ in
Figure 1) that acts on the transmitted optical signal by an (presumably
unwanted) {\it unitary} transformation ${\Bbb U}$, e.g. corresponding to
slowly fluctuating birefringence effects due possibly to temperature changes
or to local changes of the curvature of the optical fiber etc. If the
overall communication process implies the transmission of the signal from A
to B and then back from B to A, the spurious ${\Bbb U}$ effect can be
cancelled if the corresponding device (e.g. the optical fiber) is terminated
by a $FRM$ device. For instance, if the $FRM$ device is placed at the site B
the signal sent along the fiber from A to B and the received back results to
A\ to be unaffected by the disturbance introduced by the $PC$ device. In
summary, looking through the fiber A can only detect the effect of $FRM$ and 
{\it not} of $PC$. All this can be easily analyzed with the help of the
discussion above. Let the fluctuating effect contributed by $PC$ and
affecting the transmission from A to B be represented generally by the {\it %
unitary} ${\Bbb U}=\exp [i%
{\frac12}%
(\overrightarrow{\sigma }\bullet \overrightarrow{n})\theta ]$ where $%
\overrightarrow{n}$ is a unit vector pointing in {\it any} direction in the
spin space. The propagation back from B to A will be then represented by a
corresponding unitary ${\Bbb U}^{\prime }=\exp [-i%
{\frac12}%
(\overrightarrow{\sigma }\bullet \overrightarrow{n}^{\prime })\theta ]$
where the phase $(-\vartheta )$ accounts for the propagation reversal and
the components of the unit vector $\overrightarrow{n^{\prime }}$ in the {\it %
spin space} are connected to $\overrightarrow{n}$ as follows: $n_{1}^{\prime
}=n_{1}$, $n_{2}^{\prime }=-n_{2}$, $n_{3}^{\prime }=-n_{3}$. In other
words, by retracing the propagation back through the fiber the rotation-axis
involved in the ${\Bbb U}^{\prime }$ transformation has undergone a rotation
by an angle $180%
{{}^\circ}%
$ around the axis $\sigma _{1}$ respect to the original direction: $%
\overrightarrow{n}\rightarrow \overrightarrow{n}^{\prime }$. All this can be
explained by considering that the fluctuating optical disturbance ${\Bbb U}$%
, e.g. physically due to birefringence etc., consists of an
electrical-polarization $\overrightarrow{{\cal P}}$ effect. Since $%
\overrightarrow{{\cal P}}$ is a polar (''true'') vector, it is not
insensitive to any {\it sign change} of the spatial coordinates in {\it real 
}space, e.g. the one implied by the $FRM-${\it reflection} \cite{10}.
Furthermore, we have already shown that the input/output dynamics due to the 
$FRM-${\it reflection }implies{\it \ }precisely{\it \ }a{\it \ }rotation{\it %
\ }by an angle $180%
{{}^\circ}%
$ around{\it \ }$\sigma _{1}$ of any unit vector representative of the
input/output qubits:$\;\overrightarrow{p}\rightarrow \overrightarrow{p}%
^{\prime }$. The overall process can be now formally expressed by a
straightforward application of spin algebra leading to the transformation
affecting any qubit forwarded by A in a round trip journey through the whole
device $(PC{\Bbb +}FRM)$: ${\Bbb U}^{\prime }{\Bbb U}_{FRM}{\Bbb U}\equiv
\exp [-i%
{\frac12}%
(\overrightarrow{\sigma }\bullet \overrightarrow{n}^{\prime })\theta ]\times
(i\sigma _{1})\times \exp [i%
{\frac12}%
(\overrightarrow{\sigma }\bullet \overrightarrow{n})\theta ]$ =$i\sigma _{1}$%
=${\Bbb U}_{FRM}$, for {\it any }vector $\overrightarrow{n}$ and angle $%
\theta $ \cite{15}. In other words, the effect of {\it any} possible
round-trip {\it unitary} ${\Bbb U}-$process, e.g. introduced by any optical
fiber or by any other optical device, is exactly cancelled, i.e.
''compensated'', by this very clever {\it universal} $FRM$ technique.

The compensation effect has been experimentally checked by an ''ergodic
procedure'', i.e. by insertion of a stochastically time-varying unitary
operation ${\Bbb U}(t)$ into the mode ${\bf k}_{2}$ between the NL crystal
and the $FRM$. More precisely this device, represented by $PC$ in Figure 1,
and apt to simulate the most general propagation inside an optical fiber,
consisted of a cascade of two equal {\it Pockels Cells (Shangai Institute of
Ceramics, risetime }$\approx ${\it \ 1 nsec) }mutually rotated{\it \ }by{\it %
\ }$45%
{{}^\circ}%
$ and driven by a stochastic time-sequence of high-voltage electric pulses.
The output density matrices $\widetilde{\rho }_{k1,-k2}^{out}$ and $\rho
_{k1,-k2}^{out}$ representing the states of each photon pair after
propagation into the $PC+FRM$ device, respectively upon activation and
inactivation of the Pockels Cells, were reconstructed by a tomographic
technique. The value of the ''fidelity'' ${\cal F}\left( \rho
_{k1,-k2}^{out},\widetilde{\rho }_{k1,-k2}^{out}\right) =\left( Tr\sqrt{%
\sqrt{\widetilde{\rho }_{k1,-k2}^{out}}\rho _{k1,-k2}^{out}\sqrt{\widetilde{%
\rho }_{k1,-k2}^{out}}}\right) ^{2}$ expressing the overlapping of the two
experimentally determined quantum states has been found: ${\cal F}=\left(
99.9\pm 0.1\right) \%$. This figure shows the very high degree of
compensation attainable by the method.

We are indebted with Vladimir Buzek for stimulating and lively discussions
on $FRM-${\it reflection }and Marco Ricci for help with the tomographic
techniques. This work has been supported by the FET European Network on
Quantum Information and Communication (Contract IST-2000-29681: ATESIT), by
MIUR Cofinanziamento-2002 and by PRA-INFM\ 2002 (CLON).

\centerline{\bf Figure Captions}

\vskip 8mm

\parindent=0pt

\parskip=3mm

Figure.1. Experimental setup adopted for the characterization of the $FRM$-%
{\it reflection} process. INSET: Representation of the $FRM$ action on the
Poincare' sphere. The dotted lines (blue) represent the three
transformations due to the round-trip in the Faraday rotator (Paths $PO$ and 
$O^{\prime }P^{\prime }$) and to the mirror reflection (Path $OO^{\prime }$%
). The dashed line (red) represents the overall operation due to $FRM-${\it %
reflection}, i.e. a rotation around the $\sigma _{1}$ axis by an angle $180%
{{}^\circ}%
$.

Figure.2. ({\bf a}) Tomographic representation of the input and output
entangled states $\rho _{k1,k2}^{in}$and $\rho _{k1,-k2}^{out}$. Only the 
{\it real }parts is reported here. The maximum size of the experimentally
determined {\it imaginary} parts is respectively 0.02 and 0.05 in the
corresponding reported scales. The upper left diagram has been obtained by
direct measurements taken on the mode ${\bf k}_{2}$, i.e. by replacing $FRM$
with the apparatus terminated by $D_{2}$:\ figure 1. ({\bf b})
Reconstruction by {\it Quantum Process Tomography }of the matrix ${\bf M}%
_{FRM}$ representing the $FRM$-{\it reflection}.

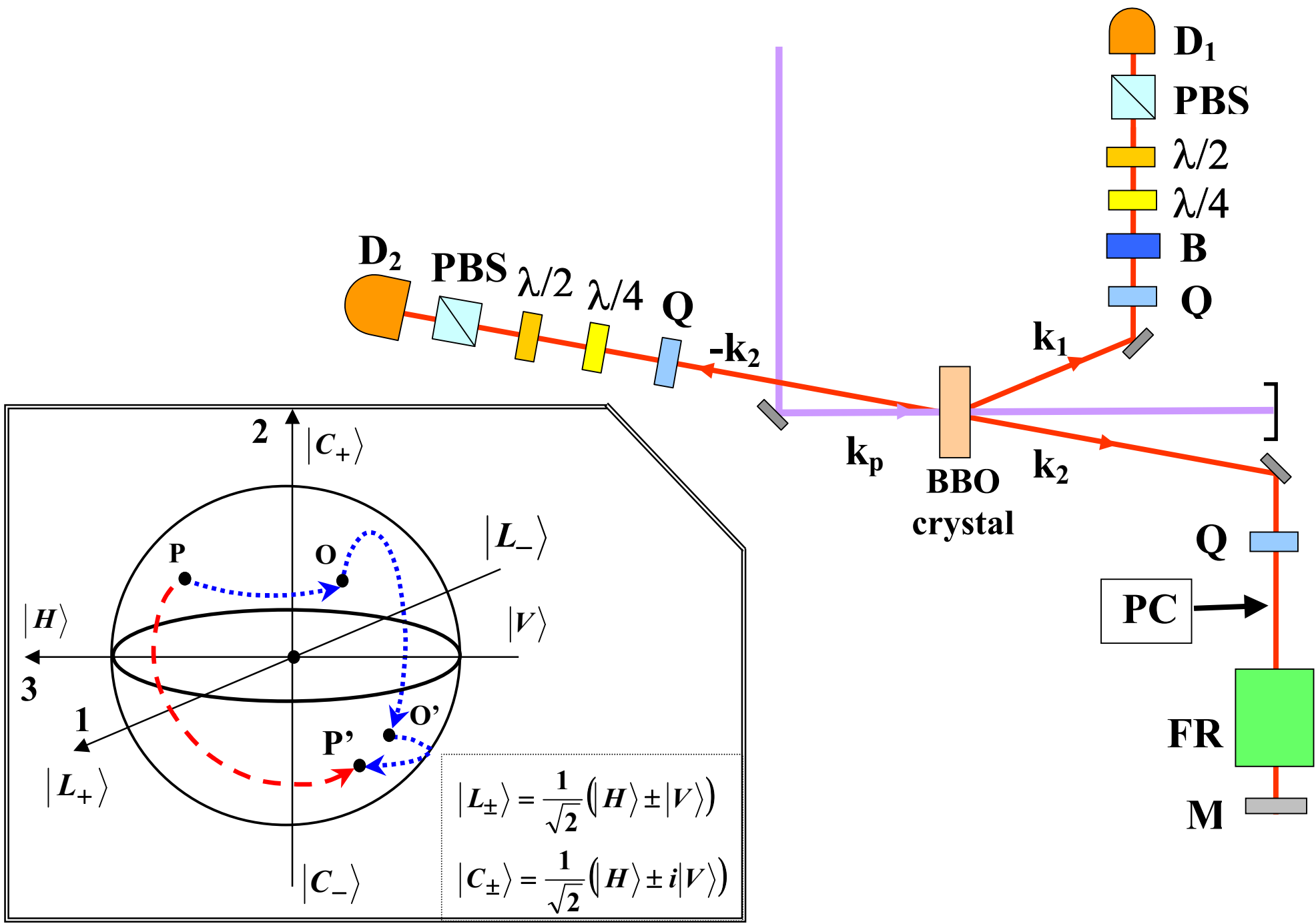
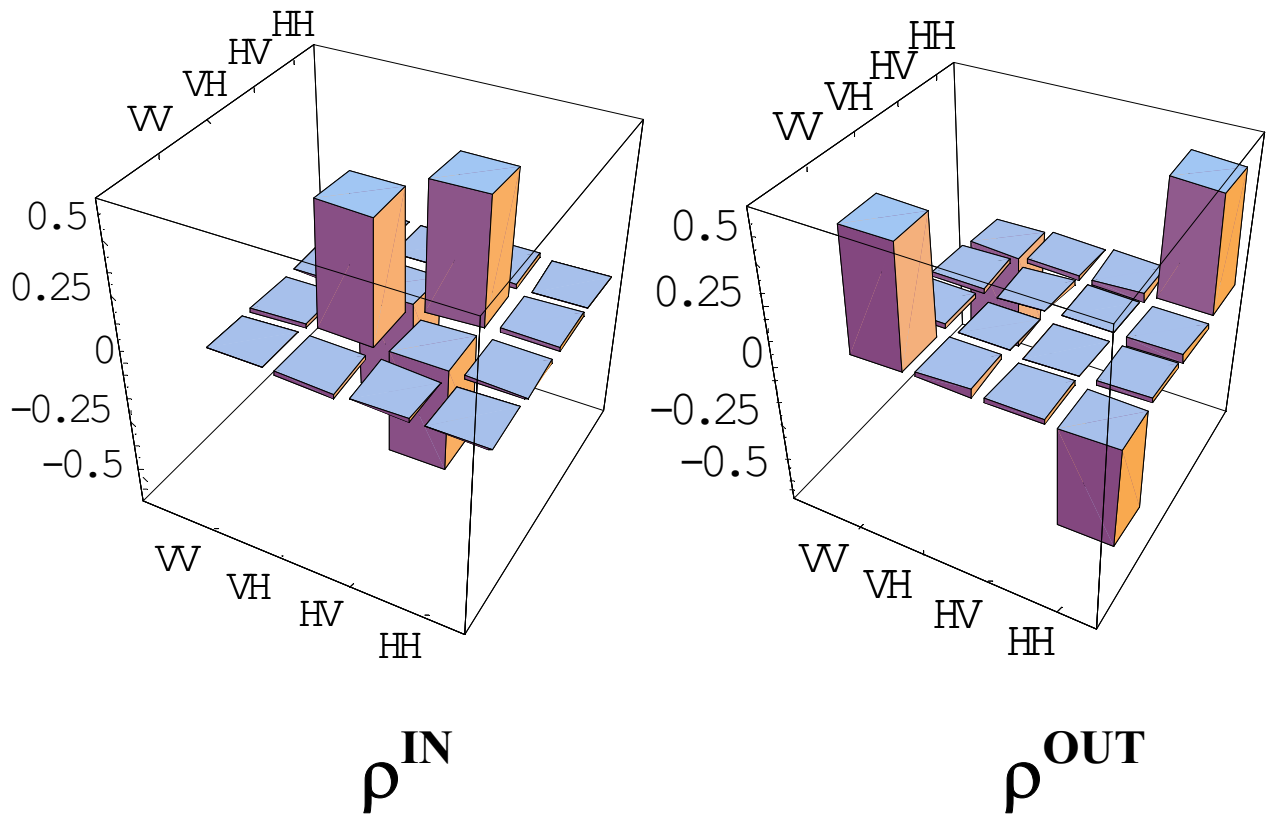
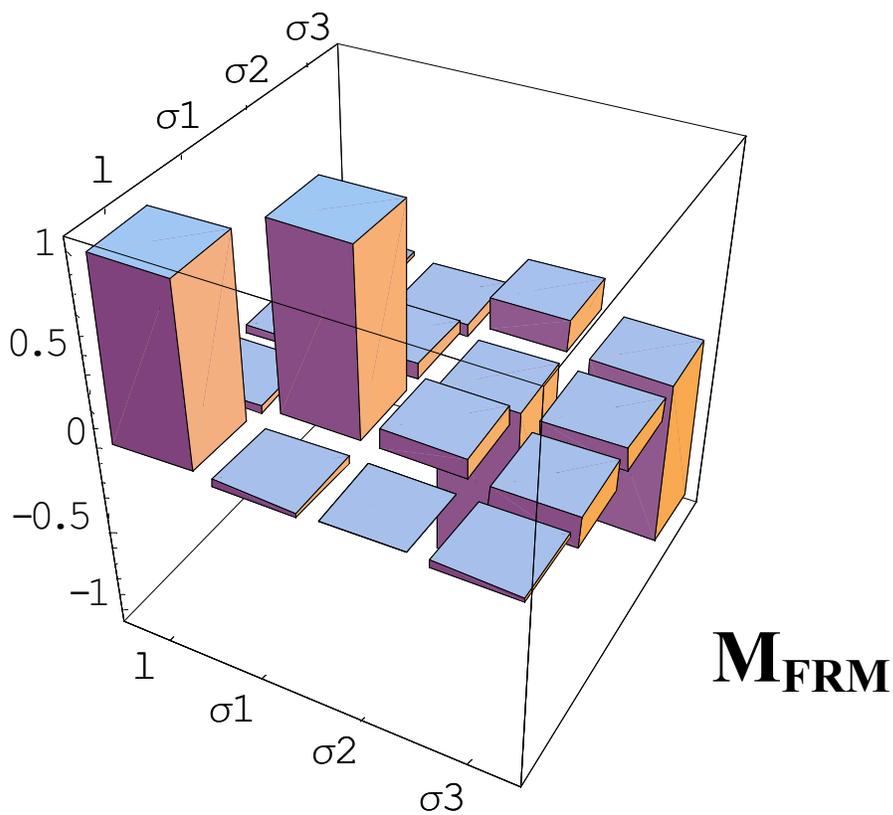
\end{document}